\documentclass{iaus}
\usepackage{graphicx}
\usepackage{enumerate}

\def\eg{{e.g.,~}}
\def\ie{{i.e.,~}}

\title[EBL from AEGIS galaxy SED-type fractions] 
{An empirical approach to the extragalactic background light from AEGIS galaxy SED-type fractions}

\author[Alberto Dom\'inguez]   
{Alberto Dom\'inguez$^{1,2,3}$
}

\affiliation{$^1$UCO/Lick Observatory, Dept. of Astronomy \& Astrophysics, University of California, Santa Cruz, CA 95064, USA\\[\affilskip]
$^2$Dept. of Physics, University of California, Santa Cruz, CA 95064, USA\\[\affilskip]$^3$Now at: Dept. of Physics \& Astronomy, University of Caliornia, Riverside, CA 92521, USA\\ email: {\tt alberto.dominguez@ucr.edu} }

\pubyear{2011} 
\volume{284}  
\pagerange{1--5}
\jname{The Spectral Energy Distribution of Galaxies}
\editors{R.J. Tuffs \&  C.C.Popescu, eds.}
\begin{document}

\maketitle

\begin{abstract}
The extragalactic background light (EBL) is of fundamental importance both for understanding the entire process of galaxy evolution and for $\gamma$-ray astronomy. However, the overall spectrum of the EBL between 0.1 and 1000~$\mu$m has never been determined directly neither from observed luminosity functions (LFs), over a wide redshift range, nor from any multiwavelength observation of galaxy spectral energy distributions (SEDs). The evolving, overall spectrum of the EBL is derived here utilizing a novel method based on observations only. It is emphasized that the local EBL seems already well constrained from the UV up to the mid-IR. Since different independent methodologies such as direct measurement, galaxy counts, $\gamma$-ray attenuation and realistic EBL modelings point towards the same EBL intensity level. A relevant contribution from Pop III stars to the local EBL seems unlikely.
\end{abstract}

\firstsection 

\section{Introduction}
\label{sec:intro}

The extragalactic background light (EBL) is the accumulated radiation in the universe from star formation process, plus a contribution from active galactic nuclei (AGNs). These photons mostly lie in the range $\sim$0.1-1000~$\mu$m. The direct measurement of the EBL is a very difficult task subject to high uncertainties. This is mainly due to the contribution of zodiacal light, some orders of magnitude larger than the EBL (\eg \cite[Hauser \& Dwek 2001]{hauser01}; \cite[Chary \& Pope 2010]{chary10}). Interestingly, it has been recently claimed by \cite{matsuoka11} the detection of the EBL free of zodiacal light. Other observational approaches set reliable lower limits on the EBL, such as measuring the integrated light from discrete extragalactic sources (\eg \cite[Madau \& Pozzetti 2000]{madau00}; \cite[Fazio et al. 2004]{fazio04}; \cite[Keenan et al. 2010]{keenan10}). On the other hand, there are phenomenological approaches in the literature that predict an overall EBL model (\ie between 0.1 and 1000~$\mu$m and for any redshift). These are basically of the four kinds described in \cite{dominguez11a} and enumerated here in Table~\ref{tab1}. Generally, any EBL modeling is built from two main quantities: one describing the galaxy density evolution over time and another one describing the overall galaxy emission (stellar component plus absorption/re-emission by dust). Table~\ref{tab1} briefly summarizes how these two main quantities are treated in the most relevant EBL modelings among the bibliography. We stress that the previous four-types classification is based upon how the different methodologies describe the galaxy density evolution.






\begin{table}
\centering
\begin{tabular}{|p{3.5cm}|p{3.5cm}|p{5cm}|}
\hline
Type of modeling \& refs. & Galaxy density evolution & \hspace{1.2cm}Galaxy emission\\
\hline
\hline
(i) Forward evolution, \eg \cite{somerville11,gilmore11} & {\bf Semi-analytical models} & {\bf Modeled}. Stellar emission: \cite{bruzual03} (BC03); Dust absorption: \cite{charlot00}; Dust re-emission: templates by \cite{rieke09}\\
\hline
(ii) Backward evolution, \eg \cite{franceschini08} & {\bf Observed} local-optical galaxy luminosity function (LF, for starburst population) and near-IR galaxy LF observed up to $z=1.4$ (for elliptical and spiral populations) & {\bf Modeled}. A few galaxy types morphologically classified based on optical images.\\
\hline
(iii) Inferred evolution, \eg \cite{finke10,kneiske10} & {\bf Parameterization} of the history of the star formation rate density of the universe & {\bf Modeled}. Stellar emission: Single bursts of solar metallicity from \cite{bruzual93} (in \cite[Kneiske \& Dole 2010]{kneiske10})/BC03 (in \cite[Finke et al. 2010]{finke10}); Dust absorption: General extinction law; Dust re-emission: Modified black bodies. AGN galaxies are not considered.\\
\hline
(iv) Observed evolution, \cite{dominguez11a} & {\bf Observed} near-IR galaxy LF up to $z=4$ & {\bf Observed}. Based on multiwavelength photometry from the UV up to MIPS~24 for $\sim 6000$ galaxies up to z=1. Consider 25 different galaxy types including AGN galaxies.\\
\hline
\end{tabular}
\caption{Classification and comparison of the main characteristics of recent EBL modelings.}
\label{tab1}
\end{table}


We consider the theoretical approach taken in \cite{somerville11,gilmore11} as complementary to our observationally motivated one to eventually reach a complete understanding of galaxy evolution. Approaches type (ii) are potentially problematic because they imply extrapolations backwards in time of local or low-redshift luminosity functions (LFs). Intrinsically different galaxy populations exist at high redshifts, which cannot be accounted for by these extrapolations. Particularly, \cite{franceschini08} use observed LFs in the near-IR from the local universe to $z=1.4$ for describing the elliptical and spiral populations, and only local for describing irregular/starbursting galaxies. They distinguish between these galaxy morphologies using images from different instruments. Different local LFs and data sets in the IR are used to constrain the mid and far-IR background. Their modelling is complex and not reproducible. Despite these particular problems, this methodology is based upon LFs, quantity directly observed and well understood unlike type (iii) models based on parameterizations of the history of the SFR density of the universe, quantity with large uncertainties and biases.

One important application of the EBL for $\gamma$-ray astronomy is to recover the unattenuated spectra of extragalactic sources. This will not be discussed in this proceeding but we refer to the interested reader to \cite[Dom\'inguez et al. 2011a,b]{dominguez11a,dominguez11b} for a discussion about this.

\section{Methodology}
\label{sec:method}

Our model is based on the rest-frame $K$-band galaxy LF found in \cite{cirasuolo10} and on multiwavelength galaxy data from the All-wavelength Extended Groth Strip International Survey (AEGIS\footnote{http://aegis.ucolick.org/}, \cite[Davis et al. 2007]{davis07}) of about 6000 galaxies in the redshift range of 0.2-1. The \cite{cirasuolo10} LF is used to count galaxies (and therefore to normalize the total EBL spectral intensity) at each redshift. The LF as well as our galaxy sample is divided into three magnitude bins according to the absolute rest-frame $K$-band magnitude, \ie faint, middle and bright. Within every magnitude bin, an SED type is statistically attached to each galaxy in the LF assuming SED-type fractions that are a function of redshift within those magnitude bins. This is estimated by fitting our AEGIS galaxy sample to the 25 galaxy-SED templates from the SWIRE\footnote{http://www.iasf-milano.inaf.it/$\sim$polletta/templates/swire$\_$templates.html} library (\cite[Polletta et al. 2007]{polletta07}). Then, luminosity densities are calculated from these magnitude bins from every galaxy population at all wavelengths, and finally all the light at all redshifts is added up to get the overall EBL spectrum.

\section{Results and conclusions}
\label{sec:results}

\begin{figure}
\centering
\includegraphics[width=10cm]{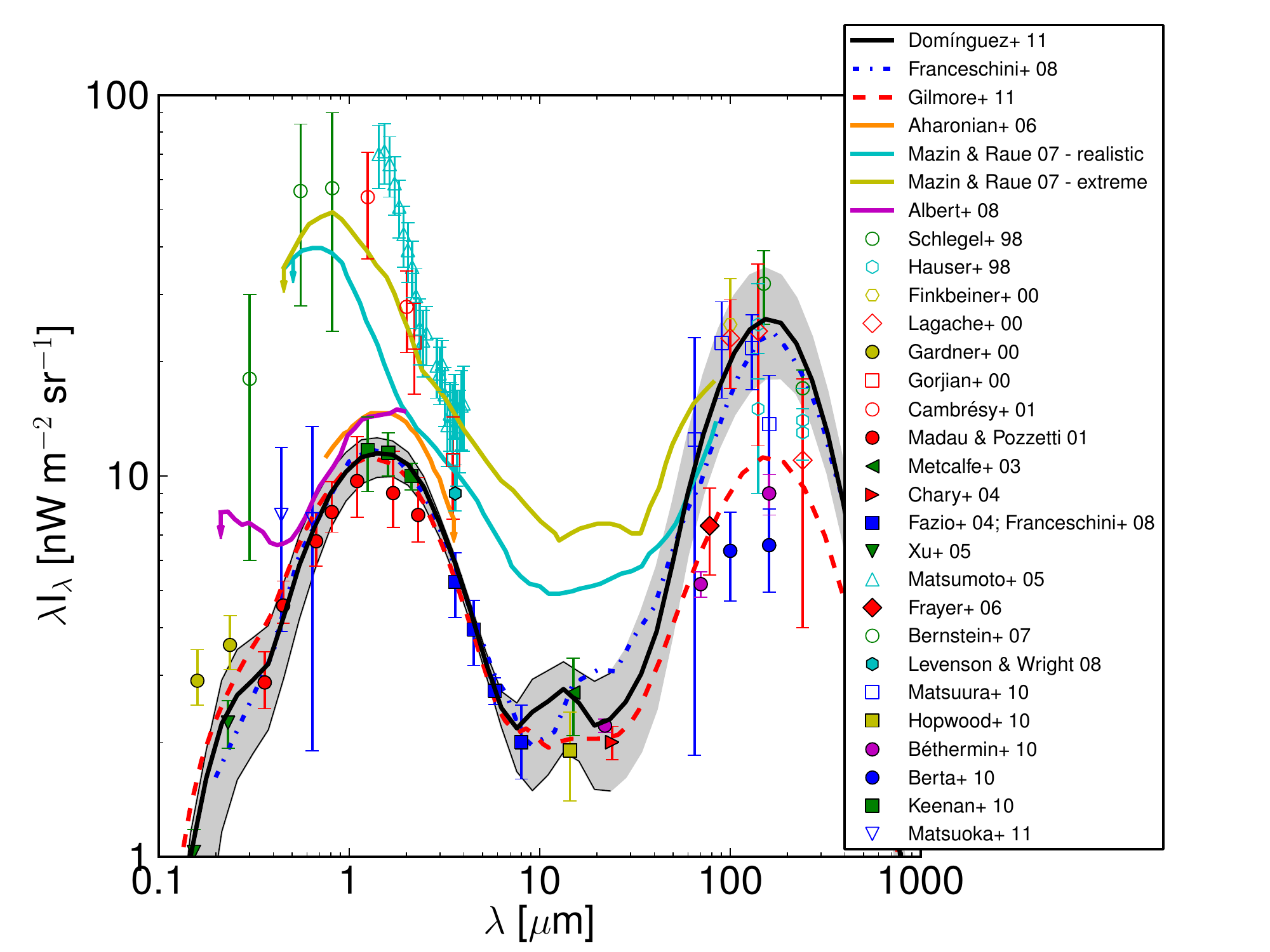}
\caption{The solid-black line is the extragalactic background light calculated from our methodology. Direct data, data from galaxy count, upper limits from $\gamma$-ray astronomy and other recent EBL modelings are shown as well (see \cite[Dom\'inguez et al. 2011a]{dominguez11a} for details).}
\label{fig:ebl}
\end{figure}

It is shown in Fig.~\ref{fig:ebl} the local EBL, with its uncertainties, compared with direct and indirect observational data, and other EBL models. Other quantities such as the EBL evolution\footnote{EBL specific intensities are publicly available at http://side.iaa.es/EBL} are discussed in \cite{dominguez11a}. Fig.~\ref{fig:ebl} suggests that the EBL coming from galaxies is already well constrained in the region from the UV up to the mid-IR but not in the far-IR. The EBL measurements free of zodiacal light in two optical bands by \cite{matsuoka11} agree with our EBL estimations. Furthermore, galaxy counts from very deep surveys taken with very sensitive instruments (\cite[Madau \& Pozzetti 2000]{madau00}; \cite[Fazio et al. 2004]{fazio04}; \cite[Keenan et al. 2010]{keenan10}) should be considered as a good estimation of the true EBL from galaxies. On the other hand, different fully independent modelings based on different methodologies and galaxy data sets such as  \cite{franceschini08,gilmore11,dominguez11a} agree in the specific intensity level of the EBL. In particular, galaxy count data are in excellent agreement with our EBL estimations. From these results, a relevant contribution from Pop III stars to the local EBL seems unlikely.

Summarizing, the best available data sets are used in our modeling (\cite[Cirasuolo et al. 2010]{cirasuolo10}'s LF and the AEGIS galaxy catalogue) observed over a wide redshift range. This model has the following main advantages over other existing EBL models: transparent methodology, reproducibility, and -very important- utilizing direct galaxy data. The galaxy evolution is directly observed in the rest-frame $K$ band up to $z = 4$. Observed galaxies up to $z = 1$ from the UV up to 24~$\mu$m with SEDs of 25 different types (from quiescent to rapidly star-forming galaxies and including AGN galaxies) are taken into account in the same observational framework. A study of the uncertainties to the model directly from the data (such as uncertainties in the Schechter parameters of the \cite{cirasuolo10} LF and the errors in the photometric catalogue) is done.

It is concluded that the EBL from galaxies seems already well constrained from UV to mid-IR wavelengths, even though uncertainties are still large in the far-IR. Furthermore, discoveries of $\gamma$-ray from distant blazars (\eg \cite[Aleksi\'c et al. 2011a,b,c]{aleksic11a,aleksic11b,aleksic11c}) support the EBL specific intensity level derived from galaxy count and recent EBL models such as \cite{gilmore11,franceschini08,dominguez11a}. We highlight that the EBL specific intensity calculated with our method is matching the lower limits from galaxy counts, which implies the highest transparency of the universe to $\gamma$-ray allowed by standard physics (see \cite[Dom\'inguez et al. 2011c]{dominguez11c}). This predicts a promising future for the new generation of imaging atmospheric Cherenkov telescopes, namely CTA.

\section*{Acknowledgements}
I thank the financial support of a Fermi grant to participate in the IAU 284 Symposium on The Spectral Energy Distribution of Galaxies.

\end{document}